 \documentclass[10pt]{article}
 \usepackage[top=1in,bottom=1in,left=1.5in,right=1.5in]{geometry}
 \usepackage{amsmath,amssymb}
  \usepackage{latexsym}
 \usepackage{graphicx}
  \usepackage{epsfig}
  \usepackage{subfigure}

\def\cF{{\cal F}}

\def\cS{{\cal S}}
\def\cX{{\cal X}}

\def\1{{\bf 1}}

\def\u{{\bf u}}
\def\v{{\bf v}}

\def\x{{\bf x}}

\def\y{{\bf y}}

\def\hs{\hspace*{\parindent}}

\def\t{^{\rm T}}

\def\R{\mathord{\mathbb R}}

\title{An Algorithm
 for Missing Value Estimation for DNA Microarray Data}
\author{Shmuel Friedland, Amir Niknejad
 \\Department of Mathematics, \\ Statistics and Computer
 Science\\ University of Illinois at Chicago
 \\ Chicago, Illinois 60607-7045\\
 friedlan@uic.edu, niknejad@math.uic.edu
 \and Mostafa Kaveh, Hossein
 Zare  \\Department of Electrical\\and Computer Engineering \\
 University of Minnesota\\Minneapolis, MN 55455\\
 \{mos, hossein\}@ece.umn.edu}

 \date {  }

\begin{document}
\maketitle

 \begin{abstract}
 Gene expression data matrices often contain missing
 expression values.  In this paper, we
 describe a new algorithm, named \emph{improved
 fixed rank approximation algorithm} (IFRAA),
 for missing values estimations of
 the large gene expression data
 matrices.
 We compare the present algorithm with the two existing and widely used methods
 for reconstructing
 missing entries for DNA microarray gene expression data: the Bayesian
 principal component analysis (BPCA) and the local least squares imputation method
 (LLS). The three algorithms were applied to four
 microarray data sets and two synthetic low-rank data matrices.
 Certain percentages
 of the elements of these data sets were randomly deleted, and the
 three algorithms were used to recover
 them.
 In conclusion IFRAA appears to be the most
 reliable and accurate approach for recovering missing DNA microarray
 gene expression data, or any other noisy data matrices that are effectively low rank.

\end{abstract}


%

\section{Introduction}
DNA microarrays  are used as a tool for analyzing information in
gene expression data over a broad range of biological applications
such as cancer classification \cite{GSL}, cancer prognosis
\cite{SRP} and identifications of cell cycle-regulated genes of
yeast \cite{Spel}. During the laboratory process, some spots on the
array may be missing due to various factors (for example, machine
error.) Because it is often very costly and time consuming to repeat
the experiment, molecular biologists, statisticians, and computer
scientists have made attempts to recover the missing gene
expressions by some ad-hoc or systematic methods.

Microarray gene expression data is often represented as a gene
expression matrix $G=(g_{ij})_{i,j=1}^{n,m}$ with $n$ rows, which
correspond to genes, and $m$ columns, which correspond to
experiments.  Thus $g_{ij}$ is the expression of the gene $i$ in
the $j-th$ experiment.  Typically $n$ is much larger than $m$. In
this setting, the analysis of missing gene expressions on the
array would translate to recovering missing entries in the gene
expression matrix values.

 In the last six years there have been at least six published papers
 in the literature discussing the problems of missing gene
 expression data
 and algorithms to recover them:
 the Bayesian principal component analysis (BPCA) \cite{Oba};
 the fixed rank approximation algorithm  (FRAA) \cite{FNC};
 the weighted K-nearest neighbors (KNNimpute) \cite{TCSBHTBA};
 the least squares principal (LSP) \cite{BDJ};
 the local least squares imputation method (LLS) \cite {KGH};
 the projection onto convex sets methods (POCS) \cite{GLY}.

 The purpose of this paper is to introduce the improved fixed rank approximation algorithm (IFRAA). We compare IFRAA with BPCA and LLS,
 since the software programs for implementing these methods are easily
 available. We have omitted comparison with KNNimpute, since the
 simulations of \cite{Oba} and \cite{KGH} show that BPCA and
 LLS are superior to KNNimpute.

 KNNimpute and LLS
 are local methods, which use similarity structure of the data to
 impute the missing values.
 KNNimpute uses the weighted averages of the $K$-nearest uncorrupted neighbors.
 LLS has two versions to find similar genes whose expressions
 are not corrupted:
 the $L_2$-norm
 and the Pearson's correlation coefficients.  After a group of
 similar genes $C$ are identified, the missing values of the gene are
 obtained using least squares applied to the group $C$.
 In these two methods, the recovery of
 missing data is done independently, i.e. the estimation of each
 missing entry does not influence the estimation of the other
 missing entries.

 BPCA is a global method
 consisting of three components.  First, principal component regression,
 which is basically a low rank approximation of the data set is performed.
 Second,
 Bayesian estimation, which assumes that the residual error
 and the projection of each gene on principal components behave
 as normal independent random variables with unknown parameters, is
 carried out. Third, Bayesian estimation follows by
 iterations based on the expectation-maximization (EM) of the unknown
  Bayesian parameters.

 IFRAA is a combination of FRAA, developed in
 \cite{FNC}, and a good clustering
 algorithm.  One first applies FRAA, whose description is below,
 to complete the missing data.  Then one applies a clustering
 algorithm to group the data to a small number of clusters of data with similar characteristics.
In each cluster FRAA is applied again to update
 the estimated values of missing entries in the cluster.

 FRAA is a global method
 which finds the values of the missing entries
 of the gene expression matrix $G$, such that
 the obtained $G$ minimizes the objective function $f_l(X)$.
 Here $f_l(X)$ is
 the sum of the squares of all
 but the first $l$ singular values of an $n\times m$
 matrix $X$.  The minimum of $f_l(X)$ is considered
 on the set $\cX$, which is the set
 of all possible
 choices of matrices $X=(x_{ij})_{i,j=1}^{n,m}$, such that $x_{ij}=g_{ij}$
 if the entry $g_{ij}$ is known.
 The completion matrix $G$ is computed iteratively,
 by a local minimization of $f_l(X)$ on $\cX$.

 The estimation of missing
 entries in FRAA is done simultaneously, i.e., the estimation of one
 missing entry influences the estimation of the other missing
 entries.

\section{Mathematical descriptions of FRAA and IFRAA}

 Let $G$ be the $n\times m$ gene expression matrix, where $n\ge m$.
 Assume first that $G$ does not have missing entries.
 Recall the \emph{singular value decomposition} of $G:=U\Sigma V\t$,
 called SVD, \cite{GV}.
 Let $\sigma_1\ge \sigma_2\ge\ldots \ge\sigma_m\ge 0$ be
 the $m$ singular values of $G$, which are the nonnegative roots
 of the eigenvalues of $G\t G$.  Let $\u_1,\ldots,\u_m\in\R^n$ and
 $\v_1,\ldots,\v_m\in\R^m$ be the column orthonormal eigenvectors
 of $G G\t$ and $G\t G$ corresponding to the eigenvalues
 $\sigma_1^2,\ldots,\sigma_m^2$ respectively.
 $\u_1,\ldots,\u_m$ and
 $\v_1,\ldots,\v_m$ are called
 the left and the right orthonormal singular
 column vectors of $G$. Then $U$ is $n\times m$ matrix, with
 the columns $\u_1,\ldots \u_m$, $V$ is an $m\times m$
 matrix, with columns $\v_1,\ldots,\v_m$, and $\Sigma$
 is the diagonal $m\times m$ matrix, with $\sigma_1,\ldots,\sigma_m$
 on the main diagonal.
 Thus $G=\sum_{i=1}^m \sigma_i \u_i\v_i\t$.
 $\u_1,\ldots,\u_m$ and $\v_1,\ldots,\v_m$ are
 the principal directions of the matrices $G G\t$ and $G\t G$
 respectively.  The rank $r$ of $G$ is equal to the number of
 positive singular values of $G$.  For each $1\le l\le r$, the
 matrix $G_l:=\sum_{i=1}^l \sigma_i\u_i\v_i\t$ is the best $n\times
 m$ approximation matrix of rank $l$.  That is if $A$ is any
 $n\times m$ matrix of rank $l$ at most, than $||G-A||_{\cF}\ge
 ||G-G_l||_{\cF}$.  ($||G||_{\cF}$ is the Euclidean norm of $G$
 viewed as a vector with $nm$ coordinates.)  An integer $l\in [1,r]$ is
 called the \emph{effective rank} of $G$, if $l$ is the smallest
 integer for which
 $\frac{\sigma_{l+1}}{\sigma_l}$ is much smaller than $1$.
 Then $G_l$ is called the \emph{filtered} $G$, and $G_l$ can be viewed
 as the noise reduction of $G$.

 In microarray analysis of the gene expression matrix $G$,
 the vectors $\u_1,\ldots,\u_m$ are called \emph{eigengenes}, the
 vectors $\v_1,\ldots,\v_m$ are called \emph{eigenarrays} and
 $\sigma_1,\ldots,\sigma_m$ are called \emph{eigenexpressions}. The
 effective rank $l$ of $G$ can be viewed as the number of different biological
 functions of $n$ genes observed in $m$ experiments. The
 eigenarrays $\v_1,\ldots,\v_l$ give the principal $l$ orthogonal
 directions in $\R^m$ corresponding to $\sigma_1,\ldots,\sigma_l$.
 The eigengenes $\u_1,\ldots,\u_l$ give the principal $l$
 orthogonal directions in $\R^n$ corresponding to
 $\sigma_1,\ldots,\sigma_l$.  The eigen expressions describe the
 relative significance of each bio-function.  From the data given
 in \cite{ABB}, one concludes that the number of significant
 singular values never exceeds $\frac{m}{2}$.
 The essence of the FRAA algorithm is
 based on this observation.

 Computationally, one brings $G$ to an upper bidiagonal matrix $A$ using
 Householder matrices.  Then one applies implicitly the QR
 algorithm to $A\t A$ to find the positive eigenvalues
 $\sigma_1^2,...,\sigma_r^2$ and the corresponding orthonormal
 eigenvectors $\v_1,...,\v_r$ of the matrix $G\t G$ \cite{GV}.

 Assume now that $G$ is the gene expression matrix with missing
 data.  We can estimate the effective rank of $G$
 by computing the effective rank of the submatrix $\hat n \times
 m$, corresponding to all genes with uncorrupted entries
 \cite[\S2 ]{FNC}.  Let $l$ be our estimate for the effective
 rank of the completed gene expression matrix.
 Denote by $\cX$
 the set of all $n\times m$ matrices whose entries coincide
 with the uncorrupted entries of $G$.  Thus $\cX$ is the set of
 all possible completion of the corrupted gene matrix $G$.
 FRAA completes the missing values of $G$ by finding the minimum
 to the following optimization problem:
 \begin{equation}\label{prob2}
 \min_{X\in \cX} \sum_{i=l+1}^m \sigma_i(X)^2= \sum_{i=l+1}^m
 \sigma_i(G^*)^2, \; \textrm{where } G^*\in\cX.
 \end{equation}
 Ideally, $G^*$ is the completion of the gene matrix expression
 with missing values.
 In practice, FRAA uses the following iterative
 procedure:

 $ $

 {\bf Fixed Rank Approximation Algorithm:} {\it
 Let $G_p\in \cX$ be the $p^{th}$ approximation to a
 solution of optimization problem (\ref{prob2}).  Let $A_p:=G_p\t G_p$ and find an
 orthonormal set of eigenvectors for $A_p$,
 $\v_{p,1},...,\v_{p,m}$.  Then $G_{p+1}$ is a solution to the
 following minimum of a convex nonnegative quadratic function}
 $ \min_{X\in \cX} \sum_{q=l+1}^m \\(X\v_{p,q})\t (X\v_{p,q})$.

 $ $

 The flow chart of this algorithm can be given as:

 $ $

 \framebox{\parbox[t]{5.1in}{

 \textbf{Fixed Rank Approximation Algorithm (FRAA)}

 \textbf{Input:}  integers $m,n,L,iter$, the locations of
 non-missing entries $\cS$,  initial approximation
 $G_0$ of $n\times m$ matrix $G$.

 \noindent
 \textbf{Output:} an approximation $G_{iter}$ of $G$.

 \noindent
 \textbf{for} $p=0$ \textbf{to} $iter-1$

 - Compute $A_p:=G_p\t G_p$ and find an
 orthonormal set of eigenvectors for $A_p$,
 $\v_{p,1},...,\v_{p,m}$.

 - $G_{p+1}$ is a solution to the minimum problem (\ref{prob2}) with $L=l$.

 }}

 $ $

 In each step of the algorithm we decrease the value of $f_l(X)$:
 $f_l(G_p) \ge f_l(G_{p+1})$.
 Hence the sequence $G_p, p=1,\ldots$
 converges to a critical point $\tilde
 G$.  Thus FRAA gives a good approximation of $\tilde G$.
 In many simulations we had we confirmed that $\tilde G=G^*$.

 Consider the following inverse
 eigenvalue problem (IEP):  {\it Find the values of the missing entries
 of $G$ such that the nonnegative definite matrix $G\t G$
 will have $m-l$ smallest eigenvalues equal to zero.}
 IEP appear often in engineering.  See \cite{FNO}
 for examples of IEP and a number of good algorithms to solve
 these problems.  In fact,
 FRAA is based on one of the algorithms for the inverse
 eigenvalue problems discussed in \cite{FNO}.

 As pointed out in \cite{FNC} FRAA is a robust algorithm
 which performs good, but not as well as KNNimpute.
 The reason of the superiority of KNNimpute lies in fact
 that it reconstruct the missing values of each gene from
 similar genes.  IFRAA discussed here overcomes
 this disadvantage.

 IFRAA works as follows.  First we use FRAA to find a completion
 $G$.  Then we use a cluster algorithm,  (we used K-means
 by repeating and refining the cluster size),
 to find a reasonable number of clusters of similar
 genes.  Presumably each cluster is a relatively smaller matrix
 having an effective low rank.
 For each cluster of genes we apply FRAA separately to recover
 the missing entries in this cluster.  It turns out that this
 modification results in a very efficient algorithm for
 reconstructing the missing values of the gene expression
 matrix.

 We also note that IFRAA performs best in
 reconstructing missing values of $n\times m$ matrices, which
 have low effective ranks.  \emph{These results suggest that IFRAA has a potential
 for being an effective algorithm
 to recover blurred spots in digital images}.

 \section{Results}

For comparison of different imputation algorithms, six different
types of data sets were used, consisting of four microarray gene
expression data and two randomly generated synthetic data. Two data
sets of microarray  were obtained from studies for the
identification of cell-cycle regulated genes in yeast
(\emph{Saccharomyces cerevisiae}) \cite{Spel}. The first gene
expression data set is a complete matrix of $5986$ genes and 14
experiments based on the Elutriation data set in \cite{Spel}. The
second microarray data set is based on Cdc15 data set in
\cite{Spel}, which contains 5611 genes and 24 experiments. Two other
yeast data sets obtained from
 "http:// sgdlite.princeton.edu\/download\/yeast\_datasets". The
 Evolution data set has been studied in \cite{Ferea} and
 Calcineurin data set has been studied in \cite{Yoshi}.
Two synthetic data set was randomly generated matrices of size $2000
\times 20$ and ranks $2$ and $8$ respectively.

 To assess the performance of missing value estimation
 methods, we performed the following simulations.
 On the first two microarray data sets and on the synthetic data we deleted
 randomly $1\%$, $5\%$, $10\%$, $15\%$
 and $20\%$ of the entries from the complete matrix $C$.
 Then we estimated
 the various completions of the missing values by BPCA,
 IFRAA and LLS.
 We set the K-value parameter
 (number of similar genes) such that there was no increase in performance of the LLS by increasing k.

 We
 used a normalized root mean square error (NRMSE) as a metric for
 comparison. If $C$ represents the complete matrix and $\hat{C}$
 represents the completed matrix using an estimate to the corrupted entries in $C$,
 then the
 root mean square error (RMSE) is $\frac{\parallel D
 \parallel_{\cF}}
 {\sqrt{N\times M}}$, where $D=C-\hat{C}$. We normalized the
 root mean square error by dividing RMSE by the average value of the
 entries in $C$.

 In IFRAA the parameter $L$,
 which is the number of
 significant singular values plus $1$, was chosen by comparison of ratio
 of two consequent singular values. We observed that this parameter appeared to
 be equal to 2 or 3 depending on data set and may differ
 for each small block of data (cluster).
 The initial guess for the missing entries in each gene was chosen to be the row
 average of its corresponding row.

 Figure 1 depicts the comparison of BPCA, IFRAA and LLS
 for Elutriation data set in \cite{Spel}.
 We break the whole gene expression matrix
 by clustering the data into groups of genes, which form matrices with
 effective low ranks.
 We applied FRAA on each group. The graph is the average over
 25 runs, and as can be seen for this data set IFRAA performed the best,
 BPCA and LLS have very close performance with significant gap
 with IFRAA.

 Figure 2 depicts the comparison of BPCA, and LLS for
 Cdc15 data set in \cite{Spel} which contains 5611 genes and 24
 experiments.
 In this case IFRAA again performed the best and LLSimpute performed slightly
 better than BPCA.

 The performance of the BCPA, IFRAA and LLS
 algorithms depends on the unknown distribution of missing position of the
 entries.
 To study this issue we applied all
 methods on the original data sets containing missing values.
 Since NRMS error could not be calculated for these
 actual missing values, we transferred the missing value positions from the
 original data to corresponding
 positions
 in the complete data derived from the original data set before applying the algorithm.
 By doing this the
 distribution of missing value positions in complete data set is almost
 unchanged from the actual distribution.
 The result is illustrated in Table I for three data sets including
 the original data set of Cdc15 which contains $\%0.7$ missing values ($\%0.81$
 missing in complete data), Evolution data set \cite{Ferea} which
 contains $\%8.457$ missing values ($\%9.1$
 missing in complete data) and Calcineurin  data set \cite{Yoshi}
 which contains $\%3.2$ missing values ($\%3.68$
 missing in complete data). This result
 again confirms the superiority of the
 IFRAA for the actual microarray data missing value
 estimation.

 The random matrices of order $2000\times 20$ and of ranks $k=2,8$
 appearing in
 Figures 3 and 4 were generated as follows.  One generates
 $2k$ random column vectors $\x_1,\ldots,\x_k\in \R^{2000}, \y_1,\ldots,
 \y_k\in \R^{20}$, where the entries of these vectors are
 chosen according to an uniform distribution. Then $C=\sum_{i=1}^k \x_i\y_i\t$.

 Figure 3 represents the comparisons of BPCA, IFRAA and LLS
 for $2000 \times 20$ random matrix of rank $2$.
 The performance of the three algorithms is excellent
 for $1\%$ of missing data.
 The performance of LLS constantly deteriorates with the increase percentage
 of missing data.  The performance of BPCA deteriorated with
 the increase percentage of missing data, but less than LLS.
 IFRAA performed outstandingly.

 Figure 4 represents the comparisons of BPCA, IFRAA and LLS
 for $2000 \times 20$ random matrix of rank $8$.
 The performance of LLS is the same as in Figure 3.
 BPCA and IFRAA performed extremely well.  IFRAA slightly
 outperformed BPCA in particular in the case with $20\%$
 of missing data.

\begin{table}
 \caption{Comparison of NRMSE for three methods: IFRAA,
 LLS and BPCA for actual missing values distribution
 for three gene expression data sets with different percentages of missing
 values.}

 \begin{tabular}{|r|l|l|l|l|}
\hline

Data sets&IFRAA &  LLS  & BPCA\\
 \hline
 Cdc15 data set \%0.81 missing  & 0.0175& 0.0200 & 0.0216\\
\hline Evolution data set \%9.16 &0.0703 & 0.0969 & 0.1247 \\
\hline Calcineurin data set \%3.68 & 0.0421 &0.0445 & 0.0453\\

\hline
\end{tabular}

\end{table}


 \section{Conclusions}

 This paper describes the improved fixed rank approximation algorithm
 (IFRAA), a local-global algorithm which exploits the local similarity in data.
 We compared IFRAA to the Bayesian principal component analysis (BPCA)
 and the local least squares imputation method (LLS).
 We applied the three algorithms to several data sets.
 We corrupted, at random, certain
 percentages of these data sets and let the three algorithms BPCA,
 IFRAA and LLS
 recover them.
 We also applied the three algorithms on real gene expression data sets while keeping
 the distribution of missing values unchanged.

 We found that
 IFRAA performed better than BPCA and LLS for
 actual microarray missing value estimation.
 In addition we observed that for microarray data sets LLS performed slightly
 better than BPCA.

 We also applied
 three algorithms on synthetic data sets, which were random
 $2000\times 20$ matrices of ranks $2$ and $8$.  We again
 corrupted at random certain percentages of these data sets.
 IFRAA and BPCA were able to recover the data quite
 well, where IFRAA slightly outperformed BPCA, in particular
 in the case with of higher percentage of missing data.
 The performance of LLS deteriorated gradually with
 increasing percentage of missing entries.

 In conclusion IFRAA appears to be the most reliable method for recovering
 missing values in DNA
 microarray gene expression data.  IFRAA was also the best to recover
 missing values in synthetic data, corresponding to a data matrix with an effectively low-rank.
 \emph{These results suggest that IFRAA has a potential
 for being an effective algorithm
 to recover blurred spots in digital images}.

\begin{figure}
\centering \scalebox{.5}{\includegraphics{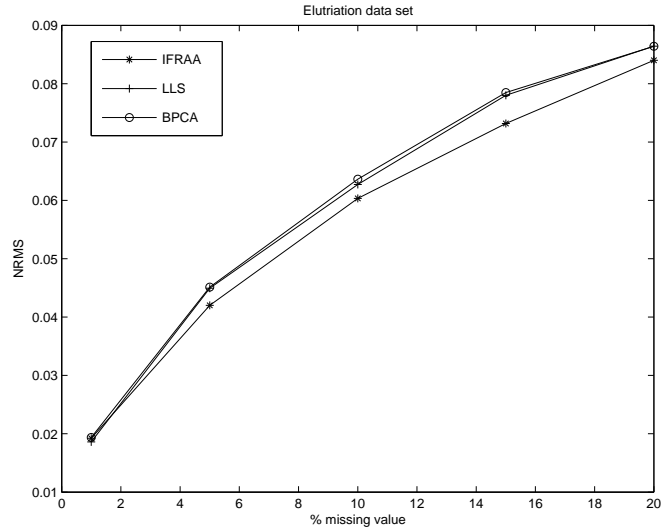}}
 \caption{Comparison of NRMSE against percent of missing entries
 for three methods: IFRAA, BPCA and LLS. Elutriation data set in
 \cite{Spel} with 14 samples.}\label{Fi:pic4}
\end{figure}

\begin{figure}
\centering \scalebox{.5}{\includegraphics{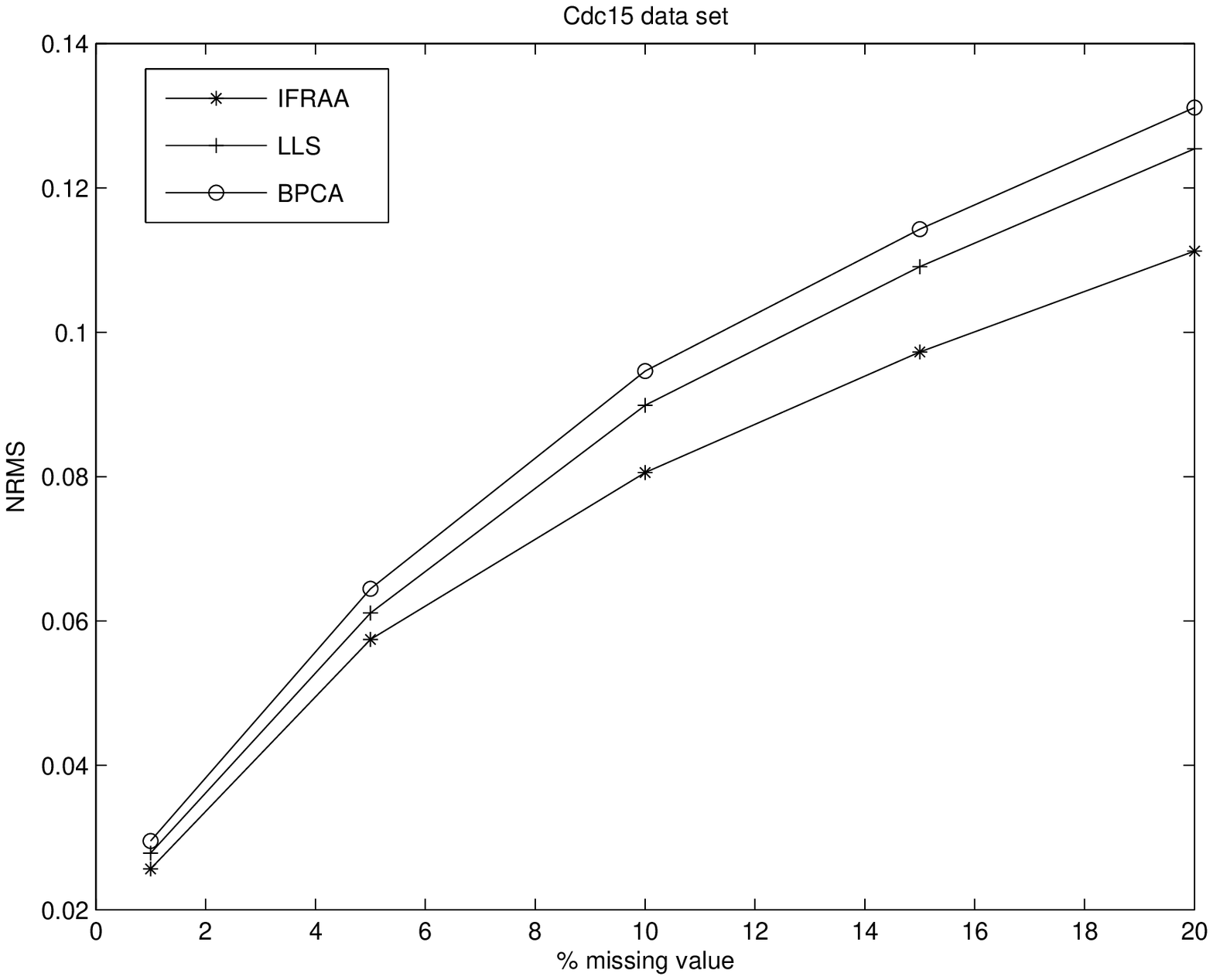}}
 \caption{Comparison of NRMSE against percent of missing entries for three methods:
 IFRAA, BPCA and LLS. Cdc15 data set in \cite{Spel} with 24 samples.}\label{Fi:pic5}
\end{figure}

\begin{figure}
\centering \scalebox{.5} {\includegraphics{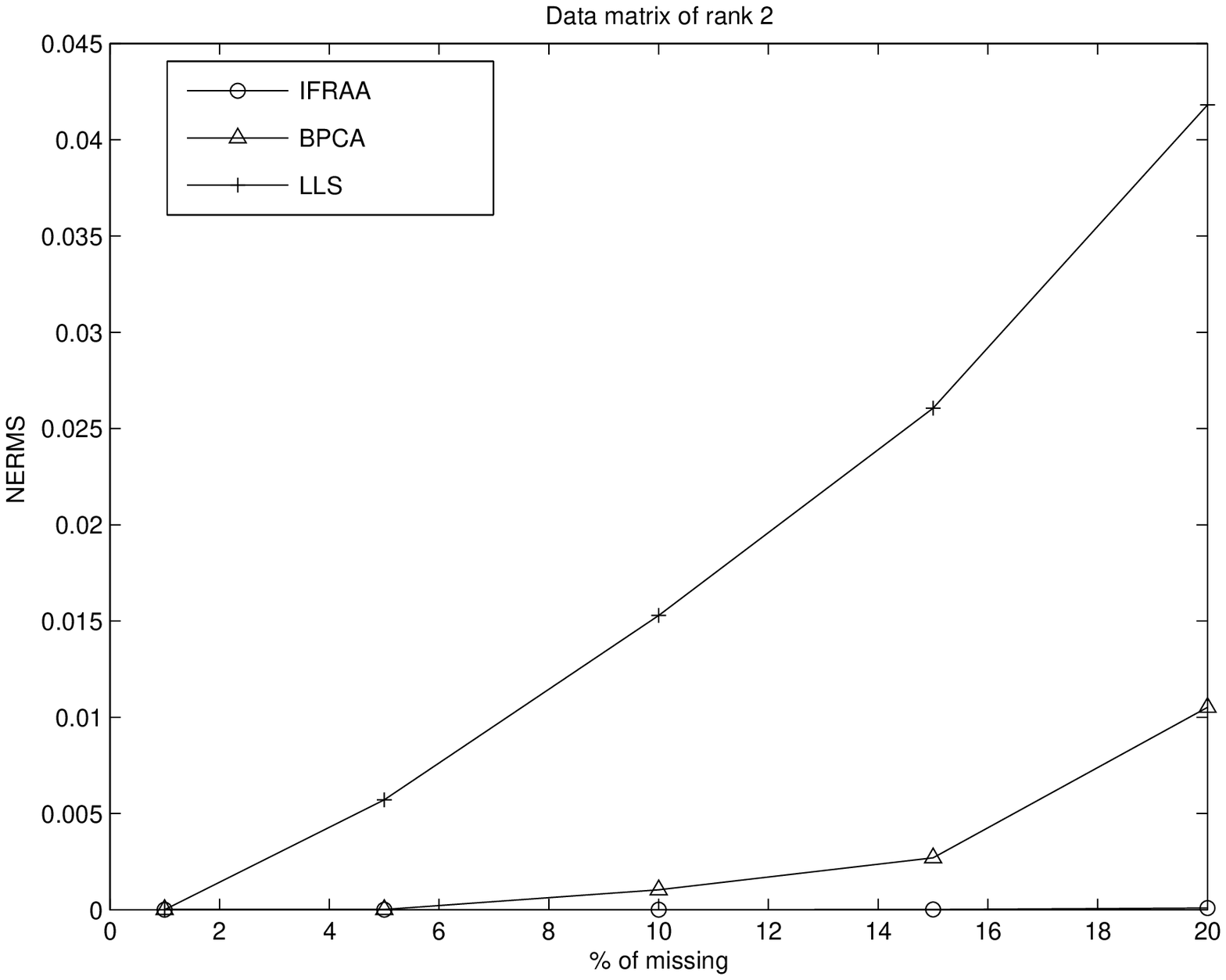}}
 \caption{Comparison of NRMSE against percent of missing entries for
 three methods:  IFRAA, BPCA and LLS. Data set was a $2000 \times 20$
 randomly generated matrix of rank 2.}\label{Fi:pic3}
\end{figure}

\begin{figure}
\centering \scalebox{.5} {\includegraphics{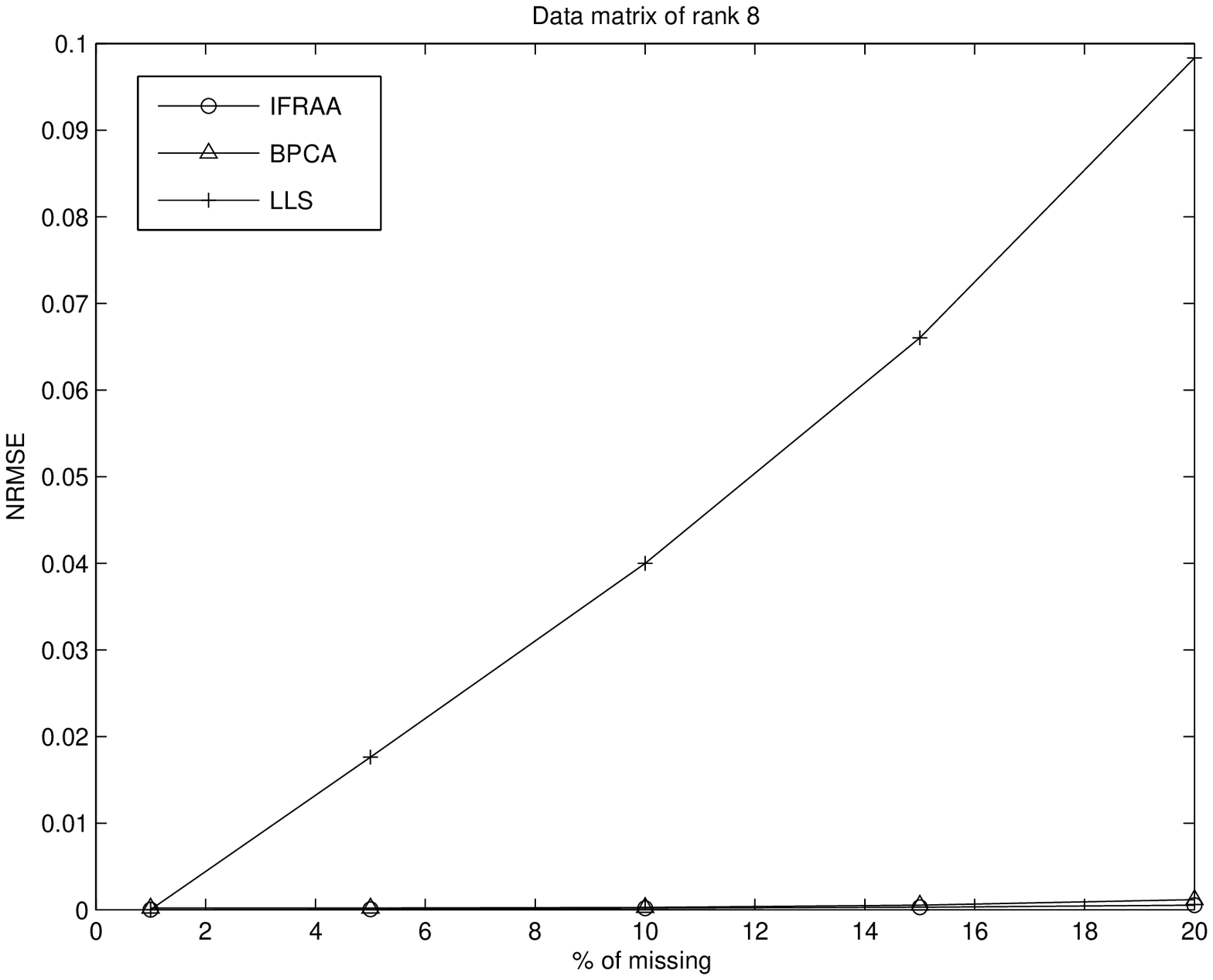}}
 \caption{Comparison of NRMSE against percent of missing entries for
 three methods:  IFRAA, BPCA and LLS. Data set was a $2000 \times 20$
 randomly generated matrix of rank 8.}\label{Fi:pic3}
\end{figure}

\end{document}